\documentclass[epj,nopacs]{svjour}
\usepackage{amsmath}
\usepackage{amssymb}
\usepackage{latexsym}
\usepackage[dvips]{graphicx}
\usepackage[english]{babel}

\def\bge{\begin{equation}}
\def\ene{\end{equation}}
\def\bgea{\begin{eqnarray}}
\def\enea{\end{eqnarray}}

\def\lemaitre{Lema$\hat{\rm i}$tre}

\def\bge{\begin{equation}}
\def\ene{\end{equation}}
\def\bgea{\begin{eqnarray}}
\def\enea{\end{eqnarray}}

\def\ls{\raise 1.5pt\hbox{$\,<\;$}\kern -10.5pt\lower3.5pt
          \hbox{$\sim$}\kern 1.5pt} %%% less or similar
\def\gs{\raise 1.5pt\hbox{$\,>\,$}\kern -9.5pt\lower3.5pt
          \hbox{$\sim$}\kern 1.5pt} %%% greater or similar

\usepackage{color}

\usepackage{ulem} % use: \sout{Striked out text}

\begin{document}
\sloppy
\title{The Anomalous 21-cm Absorption at High Redshifts}
\author{Fulvio Melia\thanks{John Woodruff Simpson Fellow.}}
\institute{Department of Physics, the Applied Math Program, and Department of Astronomy, \\
              The University of Arizona, Tucson, AZ 85721,
              \email{fmelia@email.arizona.edu}}

\authorrunning{Melia}
\titlerunning{The High-redshift 21-cm Absorption Line}

\date{\today}
%\date{March 10, 2010}
%\date{May 07, 2010}
%\date{August 8, 2018}
%\date{}

\abstract{The EDGES collaboration has reported the detection of a global 21-cm signal with a 
plateau centered at 76 MHz (i.e., redshift 17.2), with an amplitude of $500^{+200}_{-500}$ mK. 
This anomalous measurement does not comport with standard cosmology, which can only accommodate
an amplitude $\lesssim 230$ mK. Nevertheless, the line profile's redshift range
($15\lesssim z\lesssim 20$) suggests a possible link to Pop III star formation and an implied
evolution out of the `dark ages.' Given this tension with the standard model, we here examine
whether the observed 21-cm signal is instead consistent with the results of 
recent modeling based on the alternative Friedmann-Lema{\^i}tre-Robertson-Walker cosmology
known as the $R_{\rm h}=ct$ universe, showing that---in this model---the CMB radiation might have 
been rethermalized by dust ejected into the IGM by the first-generation stars at redshift 
$z\sim 16$. We find that the requirements for this process to have occurred would have 
self-consistently established an equilibrium spin temperature $T_{\rm s}\approx 3.4$ K in 
the neutral hydrogen, via the irradiation of the IGM by deep penetrating X-rays emitted at 
the termination shocks of Pop III supernova remnants. Such a dust scenario
has been strongly ruled out for the standard model, so the spin temperature ($\sim 3.3$ K)
inferred from the 21-cm absorption feature appears to be much more consistent with the 
$R_{\rm h}=ct$ profile than that implied by $\Lambda$CDM, for which adiabatic
cooling would have established a spin temperature $T_{\rm s}(z=17.2)\sim 6$ K.}
%    \PACS{{04.20.Ex},\  {95.36.+x},\  {98.80.-k},\  {98.80.Jk}}
\maketitle

% ---------------------------------------------------------------------------------
\section{Introduction}
The EDGES collaboration has recently reported the detection of a 21-cm signal
in absorption between redshifts 20 and 15, with an amplitude of 500 mK, roughly
twice the strength expected from cosmological simulations \cite{Bowman:2018}.
The most likely origin of this signal is the absorption of CMB photons by neutral
atomic hydrogen and, given that it arises from the redshift interval in which the
first generation of (Pop III) stars is believed to have formed, may have important 
implications for the early phase of cosmic structure formation.

A complication with the analysis of this signal, however, is the challenge
of making this measurement, given the very large foreground of Galactic diffuse synchrotron
emission. Though this foreground is spectrally smooth above 50 MHz, it unfortunately
requires several components for a proper modeling \cite{Bernardi:2015}. The reality
of the anomalous 21-cm absorption feature associated with the cosmic microwave
background (CMB) has therefore been called into question. Nevertheless, Bowman et al.
(2018) performed numerous hardware and processing tests to validate their detection,
demonstrating that no other astronomical or atmospheric mechanism could have produced
the observed profile. Until future observations, e.g., with the Square Kilometre
Array (SKA; ref.~\cite{Pritchard:2014}), can confirm or reject this finding, the implications
of such a global high-redshift 21-cm absorption line on cosmological models remain
somewhat ambiguous and warrant further investigation. In this paper, we shall advocate
for the reality of this anomalous signal and adopt the following essential
characteristics: the 21-cm line extends over the redshift range $15\lesssim z\lesssim 20$,
with an amplitude of $500^{+200}_{-500}$~mK at the $99\%$ confidence level. Its plateau
is centered at a frequency of $78$ MHz, corresponding to a redshift of $17.2$. The depth
of this feature is effectively $3.8\sigma$ away from the prediction of standard cosmology
($\Lambda$CDM) which, as we shall discuss shortly, cannot accommodate an amplitude
greater than $\approx 230$ mK.

Given the disparity between the measured and predicted profile of the global 21-cm
line, our principal focus in this paper will be to gauge whether this anomalous
measurement might instead offer some observational support for the alternative
Friedmann-\lemaitre-Robertson-Walker (FLRW) cosmology known as the $R_{\rm h}=ct$
universe \cite{MeliaShevchuk:2012,Melia:2020a}, which has much in common with
$\Lambda$CDM, differing with it principally via an essential constraint from general
relativity referred to as the ``zero active mass condition," i.e., an equation-of-state
$\rho+3p=0$ in the cosmic fluid, in terms of the total energy density, $\rho$,
and pressure, $p$ \cite{Melia:2019a}.

As we shall demonstrate in this paper, the gas temperature at $z=17.2$ required
to produce the observed 21-cm signal is consistent with the physical conditions
of the intergalactic medium (IGM) expected during this epoch, if the expansion
history of the Universe reflects the dynamics implied by this equation-of-state.
In \S~2, we shall summarize the status of this recent work and then, in \S~3,
describe the stellar and IGM environment expected in this model. We end with a
discussion of our results and a conclusion in \S~4.

\section{The $R_{\rm h}=ct$ Universe}
This FLRW cosmology has been under development for over a decade, and its predictions
have been tested using a broad range of data, incorporating both integrated and
differential signatures at all redshifts (for some recent developmental papers,
see refs.~\cite{Melia:2016,Melia:2017,Melia:2019a}; for a summary of the comparative
tests, see Table~2 in ref.~\cite{Melia:2018a}; a complete description of this model is
presented in ref.~\cite{Melia:2020a}). The zero active mass condition implies an
equation-of-state $p=w\rho$, with $w=(\rho_{\rm r}/3+w_{\rm de}\rho_{\rm de})/\rho=-1/3$
at all epochs, under the assumption that the pressure of matter is always negligible.
Throughout this paper, $\rho_{\rm r}$, $\rho_{\rm m}$ and $\rho_{\rm de}$ denote,
respectively, the radiation, matter and dark-energy densities, and
$p_{\rm de}=w_{\rm de}\rho_{\rm de}$. We shall write $\rho_\Lambda$ in the special
case where dark-energy is assumed to be a cosmological constant.

In this picture, the luminosity distance $d_{\rm L}$(z) and Hubble rate of
expansion $H(z)$ have very simple forms, dependent only on the parameter $H_0$:
\begin{equation}
d_{\rm L}(z) = {c\over H_0}(1+z)\ln(1+z)\;,
\end{equation}
and
\begin{equation}
H(z) = H_0(1+z)\;.
\end{equation}
The $R_{\rm h}=ct$ cosmology is flat (i.e., $k=0$), for otherwise the gravitational
radius would not equal $ct$ \cite{Melia:2018b}. Therefore, from the first Friedmann equation
\cite{MeliaShevchuk:2012}, one easily
finds that
\begin{equation}
\rho(z) = \rho_{\rm c}\,(1+z)^2\;,
\end{equation}
where
\begin{equation}
\rho_{\rm c}\equiv {3c^2\over 8\pi G}H_0^2\approx 7.7\times 10^{-9}\left({H_0\over 67.8\;
{\rm km/s/Mpc}}\right)^2\;\;{\rm erg}\;{\rm cm}^{-3}
\end{equation}
is the critical energy density. For simplicity, we scale all relevant quantities to the
{\it Planck} value of $H_0$ though, in reality, the actual $H_0$ in $R_{\rm h}=ct$ is
somewhat different. Even so, it would differ from this value by no more than
$\sim 10-15\%$ (see, e.g., ref.~\cite{MeliaMaier:2013}).

Our imposition of the zero active mass constraint on the cosmic fluid implies that the
fractional representation of the various species is modified from the evolution seen in
$\Lambda$CDM. As we shall discuss further below, this is partly the reason
why the rate of structure formation in this model, notably the halo growth rate and
stellar formation rate, differ considerably from their counterparts in the standard
model. Let us define the dark energy density as $\rho_{\rm de}=\varpi(z)\,\rho$.
The equation-of-state in $R_{\rm h}=ct$ does not tell us the value of $\varpi$, but we
know from observations of the local Universe that $\sim 30\%$ of $\rho$ at low redshifts
is comprised of (luminous and dark) matter. Thus, if $\varpi=2/3$ at $z\rightarrow 0$, the
expression for $p$ implies that $w_{\rm de}=-1/2$, precluding any possibility of
dark energy being a cosmological constant. We do not know whether this equation-of-state
changes with redshift, but if we
assume for simpicity that $w_{\rm de}$ is constant, then the early Universe must have
both radiation and dark energy, in proportions such that $\varpi=4/5$ at $z>>1$. Dark energy
is therefore always present in $R_{\rm h}=ct$, though its fractional contribution to
$\rho$ evolves slightly from $0.8$ at early times to $2/3$ locally. In concert with this
change, radiation is gradually replaced by matter, with
\begin{equation}
\rho_{\rm m}=(2-5\varpi/2)(1+z)^2\rho_{\rm c}\;,
\end{equation}
and
\begin{equation}
\rho_{\rm r}=(3\varpi/2-1)(1+z)^2\rho_{\rm c}\;.
\end{equation}

If we further reasonably assume that the radiation is always a blackbody, the
temperature evolves according to the standard free-streaming relation
\begin{equation}
T_\gamma(z) = T_0(1+z)\quad (z\lesssim z_{\rm cmb})
\end{equation}
after decoupling (at $z_{\rm cmb}$), but follows the modified form
\begin{eqnarray}
T_\gamma(z) &\approx& 17.9\,\left({3\varpi/2-1\over 0.1}\right)^{1/4}
(1+z)^{1/2}\times\nonumber\\
&\null&\hskip 0.4in\left({H_0\over 67.8\;{\rm km/s/Mpc}}\right)^{1/2}
\;{\rm K} \quad (z>>1)\quad
\end{eqnarray}
at high redshifts.

In recent work \cite{Melia:2020b}, we discussed the reasons (both theoretical
and observational) why $z_{\rm cmb}$ in this model cannot be as large (i.e., $\sim 
1080$) as in the standard model. Indeed, we may see right away from Equations~(7)
and (8) that this transition must occur, crudely, at the redshift where these two
expressions overlap, i.e., at $z_{\rm cmb}\lesssim 42$. More quantitatively,
{\it Planck} \cite{Planck:2016} has identified an angular scale
$\theta_{\rm s}=(0.596724\pm 0.00038)^\circ$ on the last scattering surface
(LSS), interpreted as the acoustic horizon---the maximum distance traveled by
sound waves in the early universe (see, e.g.,
refs.~\cite{PeeblesYu:1970,White:1994,Hu:1995,Seo:2005}). The comoving distance
$r_{\rm s}$ corresponding to this scale is thought to have remained constant
thereafter.

A characteristic scale has also been seen in the correlation function of galaxies
and the Ly-$\alpha$ forest
\cite{Meiksin:1999,Seo:2005,Jeong:2006,Crocce:2006,Eisenstein:2007,Nishimichi:2007,Seo:2010,Font-Ribera:2014,Delubac:2015,Alam:2017}. Interpreting the peak seen in large galaxy surveys as the
subsequent manifestation of the acoustic horizon, the baryon acoustic oscillation
(BAO) comoving scale in the context of $R_{\rm h}=ct$ has been determined
to have the value $r_{\rm BAO}=131.1\pm4.3$ Mpc \cite{MeliaLopez:2017,Melia:2018a},
and it is straightforward to show from Equation~(1) that
\begin{equation}
\ln(1+z_{\rm cmb})={r_{\rm BAO}\over R_{\rm h}(t_0)\,\theta_{\rm s}}\;,
\end{equation}
where $R_{\rm h}(t_0)=c/H_0$ is the gravitational (or Hubble) radius today.
Taking the measurement errors into account, we therefore infer that (in the
$R_{\rm h}=ct$ cosmology),
\begin{equation}
z_{\rm cmb}=16.05^{+2.4}_{-2.0}\;,
\end{equation}
corresponding to a cosmic time $t_{\rm cmb}\approx 849$ Myr. Matter could
therefore not have been ionized at this redshift, so the recombination
picture for the origin of the CMB could not work in this model. A recent
analysis \cite{Melia:2020b} of the dust injected into the IGM by
Pop~III stars instead suggests that the primordial radiation field would have
been rethermalized by grains at $z_{\rm cmb}\sim 16$ to produce the CMB we
see today. In this picture, the CMB anisotropies therefore reflect the large-scale
structure seen just prior to the onset of reionization at $z\sim 15$.

Additional evidence for the value of $z_{\rm cmb}$ in Equation~(10) is provided
by the CMB angular correlation function \cite{MeliaLopez:2018}.
The lack of large-angle correlations in the CMB temperature fluctuations
conflicts with the predictions of inflationary $\Lambda$CDM at a high level
of confidence ($\gtrsim 3\sigma$; ref.~\cite{Copi:2015}). Following a careful analysis
of the latest {\it Planck} release, we now understand that the weakness of the
large-angle correlations is probably due to a non-zero minimum wavenumber $k_{\rm min}$
in the primordial power spectrum $P(k)$ \cite{MeliaLopez:2018}. Its
relevance to the subject of this paper is that, if real, the inferred value
of $k_{\rm min}$, in the context of $R_{\rm h}=ct$, corresponds to a redshift
$z_{\rm cmb} = 17.05^{+8}_{-5}$---a compelling confirmation of the estimate
shown in Equation~(10), based on an entirely different argument.

But is it really possible to even contemplate a dusty origin for the CMB, given
that all of the evidence today appears to overwhelmingly favour recombination
at $z\sim 1080$? In the next section, we shall summarize the extensive analysis
carried out to address this far-fetched idea. As it turns out, such a model was
actually seriously considered in the late 1900's before its evident flaws made
it redundant. But we now know that, while a dust model for the CMB is untenable
in $\Lambda$CDM, it may still be viable in other cosmologies. As it turns out, dust
could still play a critical role in forming the microwave background if
$R_{\rm h}=ct$ is the correct model. And as we shall see, this scenario also
self-consistently accounts for the 21-cm signal measured by the EDGES collaboration,
which (subject to confirmation by future measurements) is otherwise
not consistent with the standard model.

\section{Reassessing Dust's Role in Forming the CMB}
Placing the last scattering surface (LSS) at $z_{\rm cmb}\sim 16$ may seem to be at odds with
many kinds of observation, but this is only true in the case of $\Lambda$CDM. As shown in
ref.~\cite{Melia:2020b}, there are at least three principal observational signatures one
may use to distinguish a CMB originating via recombination at $z\sim 1080$ from one
due to reprocessing by dust at $z\sim 16$. These are: the spectrum itself, which appears
to be a true blackbody \cite{Mather:1990}; the presence or absence of recombination
lines \cite{Rubino-Martin:2006,Rubino-Martin:2008}, and a test of whether the angular
power spectrum varies with frequency \cite{Planck:2016}.

The physical attributes required of an LSS at $z\sim 16$ in the context of $R_{\rm h}=ct$
ironically echo some of the theoretical ideas explored for a dusty origin of the CMB
several decades ago \cite{Rees:1978,Rowan:1979,Wright:1982}. Before attempting
to rescue this now defunct model in order to explain the anomalous 21-cm signal, however, it
is essential to consider whether such a proposal even makes sense based on what we know today.

The first issue is quite obvious: recombination lines ought to be present at some level
in the CMB spectrum if the standard picture is correct, whereas dust rethermalization
at $z\sim 16$ would have wiped all of them out. Extensive simulations have already
been carried out to predict the level at which we should see such lines in $\Lambda$CDM
\cite{Rubino-Martin:2006,Rubino-Martin:2008}, but unfortunately the effect of recombination lines
on the angular power spectrum is expected to be quite small. Though it may eventually be
separated from other effects with the improved sensitivities of future experiments,
there is no evidence today of recombination lines in the CMB, so this test must await
future developments.

Arguably the most convincing observation made to support the recombination scenario
was COBE's discovery \cite{Mather:1990} that the CMB's spectrum is a near perfect
blackbody. We shall not repeat the argument made in ref.~\cite{Melia:2020b}, but it
is not difficult to show that optically thick dust in thermal equilibrium with the
radiation it rethermalizes near the photosphere (at the LSS) also produces a near
perfect blackbody field, as was already suggested by some of the earlier work
\cite{Rowan:1979}. Thus, the key issue is not that dust opacity is frequency dependent
but, rather, that dust would need to reach local thermal equilibrium with the radiation.
This question hinges on how much dust was produced by Pop~III stars in the redshift
range $15\lesssim z\lesssim 20$, and we shall discuss this in \S~3.1 below.

The third issue is whether the angular power spectrum of the CMB is frequency-dependent,
as one might expect if the anisotropies in the temperature distribution across the sky vary
among surveys conducted at different wavelengths. One would not expect photospheric optical
depths to affect the observed distribution of fluctuations with Thomson scattering because
the optical depth it produces is independent of frequency. Maps made at different wavelengths
should therefore be identical in $\Lambda$CDM. But this would not be the case if the opacity
were frequency dependent, as would happen in the case of dust. Though photospheric depth effects
might not significantly change the larger fluctuations from one map to another, they could
still alter the observed anisotropy distribution on smaller scales, which would in turn produce
variations in the inferred CMB power spectrum constructed at different wavelengths. A
careful analysis of this frequency dependence \cite{Planck:2018} shows
that indeed the location of the LSS changes somewhat with frequency. Variations can
be as high as $\sim 2\%$ at $\ell\sim 400$, increasing to $\sim 5\%$ for $\ell\gtrsim 800$.
If real, these changes would argue against a Thomson scattering opacity, but would be
consistent with the dust model in $R_{\rm h}=ct$ (see ref.~\cite{Melia:2020b}) as long as the
location of the LSS were restricted to the redshift range $\sim 14-16$. This period
would have lasted $\sim 100$ Myr, after which the Pop~III and the early Pop~II
supernovae would have completely destroyed the dust, making the IGM transparent again
and initiating the epoch of reionization at $z\sim 15$. Under these constraints, a
frequency-dependent opacity would have impacted the CMB spectrum by no more than a few
percent, consistent with the current {\it Planck} observational measurements.

\subsection{Pop~III Stars in the $R_{\rm h}=ct$ Universe}
Assessing the CMB-imposed constraints on the Pop~III star formation rate and evolution
is critical to understanding whether the IGM at $15\lesssim z\lesssim 20$ could also have
produced the observed global 21-cm signal. Again, we won't repeat the detailed analysis
carried out in ref.~\cite{Melia:2020b}, but we shall here merely summarize the key findings from that
work. In order for every photon in the CMB to have been absorbed by dust prior to $z\sim 16$,
we would require the dust number density to have been $n_{\rm s}(z=16)\sim 5\times 10^{-15}
f_{\rm Z}$ cm$^{-3}$ for a bulk density of $\sim 2$ g cm$^{-3}$ of silicate grains, and
a grain radius $r_{\rm s}\sim 0.1$ micron. This restricts the IGM metallicity $f_{\rm Z}$
near the end of the Pop III star formation and evolution era to $\sim 20\%$ of the solar
value, comfortably small to avoid any obvious inconsistency with the general
perception that the bulk of today's abundances grew across subsequent phases of star formation.

In order for the dust temperature to have remained in equilibrium with the CMB during that
period, two important factors had to be satisfied. The first has to do with the average heating
and cooling rates for a given dust particle, while the second is based on the fact that each
absorption of a photon produces a quantum change in the dust particle's temperature, and is thus
strongly dependent on its size \cite{Weingartner:2001,Draine:2001}. Under the conditions
expected in the IGM at that redshift, it would have taken a $\sim 0.1\mu$m sized particle roughly
$50$ seconds to reach equilibrium at a dust temperature $T_{\rm d}\sim 46$ K, so the first
condition would have been satisfied trivially. The second issue is more constraining. The
assumption of a smooth evolution in $T_{\rm d}$ starts to break down for grains smaller than
$r_{\rm s}\sim 0.003$ $\mu$m \cite{Draine:2001}, at which point the heating starts to produce
temperature spikes. Putting these estimates together, we see that the dust model required for
consistency with the observed CMB spectrum would therefore be based on silicates with size
$\sim 0.003-0.3$ $\mu$m, perhaps even larger, though sizes much larger than $\sim 0.3$~$\mu$m
would violate our earlier estimate of $n_{\rm s}(z=16)$ and the reasonable value of the
metallicity $f_{\rm Z}\sim 0.2$.

Let us now see what these constraints have to say about the Pop III star formation rate
and evolution. Believed to have reached masses $500\;M_\odot\gtrsim M\gtrsim 21\;M_\odot$
\cite{Bromm:2004,Glover:2004}, these stars emitted copious high-energy radiation
that ionized the halos within which they were born. At the end of their brief
($\sim 10^6-10^7$ yr) lives, most of them exploded as SNe \cite{Heger:2003}, ejecting
heavy elements into the IGM \cite{Whalen:2008}, also producing X-rays via shock
acceleration that penetrated deeply into the IGM (more on this below). Given the above
estimates for the dust size and number density, one can infer that roughly
$9\times 10^{44}$ g Mpc$^{-3}$ (co-moving volume) of dust material had to be injected
into the IGM during this epoch ($20\gtrsim z\gtrsim 15$). Thus, adopting a typical
mass $M\sim 100\;M_\odot$, with a typical ejection fraction of $30\%$ \cite{Heger:2002},
we see that $\sim 1.5\times 10^8$ Mpc$^{-3}$ Pop III stars must have exploded as SNe
between $z=20$ and $15$ to provide the dust required to completely thermalize with the
CMB radiation. As we shall see below, this star-formation rate (SFR) is critical to understanding
how the global 21-cm signal could have been produced within the context of $R_{\rm h}=ct$.

Those familiar with current models of star formation during this epoch (see, e.g.,
refs.~\cite{Xu:2016a,Xu:2016b,Mebane:2018,Visbal:2018,Jaacks:2019,Sarmento:2019}
would realize that such an aggressive SFR would be orders of magnitude greater
than what is expected in the context of $\Lambda$CDM. Comparable simulations have not yet
been completed using $R_{\rm h}=ct$ as the background cosmology, but already we have firm
indications that large-scale structure (LSS) formation proceeded at a very different rate
in this model, due to several influential factors, including: (i) the seeding of a primordial
fluctuation spectrum \cite{Melia:2019b}; (ii) the alternative redshift dependence of the relative
abundances (Eqns.~5, 6); (iii) the different timeline versus redshift \cite{Melia:2014}; (iv)
the modified growth equation, which has now been solved semi-analytically
\cite{Yennapureddy:2018,Yennapureddy:2019,Yennapureddy:2020}.
What emerges from a comparison of LSS formation in these two
cosmologies is a significant difference in the growth rate of stellar mass at $z\gtrsim 4$,
reaching values exceeding $\sim 10^4$ at $z\gtrsim 10$. The observational evidence one may
use to distinguish between these two scenarios is still controversial and strongly debated,
but a quick inspection of figures~2--7 in ref.~\cite{Yennapureddy:2019} would reveal that
the data do not yet rule out the higher rates predicted by $R_{\rm h}=ct$. Indeed, the
currently inferred halo mass function appears to differ from what is expected at
$z\sim 10$ in $\Lambda$CDM by over 4 orders of magnitude \cite{Steinhardt:2016}.
Some additional support for the timeline in $R{\rm h}=ct$ is also provided by its
consistency with the early appearance of supermassive black holes \cite{Melia:2015},
which would otherwise pose quite a challenge to the time versus redshift relation
in the standard model.

In addition to the dust they subsequently ejected into the IGM, Pop III stars also emitted
copious amounts of UV light during their lives, and X-rays via their supernova remnants. A
typical Pop III star with mass $M\sim 100\;M_\odot$ was a blackbody emitter with radius
$R_*=3.9\,R_\odot$ and surface effective temperature $T_*=10^5$ K. It is straightforward
to estimate from this that, during their evolution, they would have bathed the IGM with
an energy density $U_{UV}\sim 4\times 10^{63}$ erg Mpc$^{-3}$. Compared to the energy
density in the CMB at that time, this would have amounted to no more than $\sim 0.5\%$
of the total radiation field, but these UV photons would have had important consequences
concerning the HI gas and spin temperature we shall be discussing shortly.

\subsection{The Gas and Spin Temperature at $16\lesssim z\lesssim 19$ in the $R_{\rm h}=ct$ Universe}
The principal effort in modeling the global 21-cm signal is directed towards understanding the
gas kinetic ($T_{\rm g}$) and HI spin temperature ($T_{\rm s}$) during the epoch ($z\sim 17.2$)
when the absorption line formed. The 21-cm line is the triplet-to-singlet hyperfine transition
of the atomic hydrogen ground state due to the coupling of the magnetic moments of the proton
and electron (for a review, see, e.g., ref.~\cite{Pritchard:2012}). For a system in thermal
equilibrium, the relative populations of the two spin levels are given by the ratio
\begin{equation}
{n_{\uparrow\uparrow}\over n_{\uparrow\downarrow}}={g_1\over g_0}e^{-\Delta E/T_{\rm s}}\;,
\end{equation}
where $\Delta E=5.9\times 10^{-6}$ eV is the energy difference between the two distinct
spin states, and $g_1/g_0=3$ is the ratio of the statistical degeneracy factors of the
two levels.

The spin temperature coupled to either the gas or CMB temperature, depending on which
physical processes dominated the excitation of the line at $15\lesssim z\lesssim 20$.
There appear to be three dominant effects: (1) the absorption of CMB photons which,
in the absence of other interactions, would imply that $T_{\rm s}=T_\gamma$; (2)
nucleon-nucleon collisions, which would couple $T_{\rm s}$ to the gas temperature
$T_g$; and (3) excitation due to the so-called Wouthuysen-Field effect \cite{Wouthuysen:1952}.
In this process, Ly-$\alpha$ photons produced, say, by the Pop III stars, excited the
ground state of atomic hydrogen to the 2P level, which then re-emited Ly-$\alpha$
photons and entered either of the two spin states, creating an asymmetric redistribution
of the electrons between the hyperfine levels. In equilibrium \cite{Hirata:2006}, one has
\begin{equation}
T_{\rm s}^{-1}={T_\gamma^{-1}+x_\alpha T_{\rm eff}^{-1}+x_cT_g^{-1}\over 1+x_\alpha+x_c}\;.
\end{equation}
In this expression, $x_c$ is the collisional coupling coefficient for H-H and H-e$^{-1}$
scatterings \cite{Zygelman:2005}. The middle term accounts for the Wouthuysen-Field effect,
with a coupling coefficient \cite{Hirata:2006}
\begin{equation}
x_\alpha=1.81\times 10^{11}(1+z)^{-1}S_\alpha J_\alpha\;,
\end{equation}
where $S_\alpha$ is a factor of order unity describing the detailed atomic physics
of the scattering mechanism, and $J_\alpha$ is the Ly-$\alpha$ background (number)
intensity in units of cm$^{-2}$ s$^{-1}$ Hz$^{-1}$ sr$^{-1}$. This process couples
$T_{\rm s}$ to an effective color temperature $T_{\rm eff}$ which, however, never
deviates from $T_g$ by more than $\sim 20\%$ (see fig.~4 in ref.~\cite{Hirata:2006}),
due to the recoils induced by repeated scatterings.

It is straightforward to estimate $J_\alpha$ from the total Pop III emissivity discussed
in the previous subsection.  One finds for the redshift range $15\lesssim z\lesssim 20$ in
$R_{\rm h}=ct$ that
\begin{equation}
J_\alpha\sim 2\times 10^{-7}\;{\rm cm}^{-2}\;{\rm s}^{-1}\;{\rm Hz}^{-1}\;{\rm sr}^{-1}\;.
\end{equation}
This value does not include the possible contribution from nascent quasars, which might
have started forming during this epoch as well, so it should be viewed as a lower limit. Thus,
the coefficient $x_\alpha$ in Equation~(12) would have been $\gtrsim 2,400$, and consequently
the spin temperature $T_{\rm s}$ would have been tightly coupled to the gas temperature
throughout the period where the global 21-cm absorption line was formed.

The 21-cm brightness temperature (relative to the CMB) averaged over the whole sky is given by
ref.~\cite{Zaldarriaga:2004}
\begin{eqnarray}
\delta T_b&=&27\;{\rm mK}\; x_{\rm HI}(1+\delta)\left({\Omega_{\rm b}\over 0.05}\right)
\left({0.33\over \Omega_{\rm m}}{1+z\over 10}\right)^{1/2}\times\nonumber\\
&\null&\hskip 0.3in\left({\partial_r v_r\over
(1+z)H(z)}\right)\left(1-{T_\gamma\over T_{\rm s}}\right)\;,\quad
\end{eqnarray}
where $x_{\rm HI}$ is the fraction of neutral hydrogen ($\approx 1$ in this regime),
$\delta$ is the fractional overdensity (which is smaller than $1$, so that $1+\delta\approx 1$),
and $\Omega_{\rm m}$ and $\Omega_{\rm b}$ are, respectively, the matter and baryonic
densities today in units of the critical density. The quantity $\partial_r v_r$ is the
velocity gradient along the line of sight and is approximately equal to $(1+z)H(z)$
within this redshift range.

The application of Equation~(15) to the observed global 21-cm signal thus implies that
$T_{\rm s}$ (and therefore $T_g$) must have been approximately $3.3$ K during this epoch
(remembering that $T_b$ is actually negative), compared to the CMB temperature
$T_\gamma=2.725(1+17.2)\approx 49.6$ K. So the key question is now ``How did the gas
reach this equilibrium at $15\lesssim z\lesssim 20$?"

Let us first consider the situation in the standard model. The conventional assumption
in $\Lambda$CDM is that, while the radiation effectively decoupled from the baryons at
$z\sim 1080$ (LSS), the particles were much less numerous and therefore remained thermalized
with the radiation for much longer \cite{Peebles:1968}. The rate of energy transfer per unit
volume between the CMB photons and free electrons may be written \cite{Weymann:1965}
\begin{equation}
{d\epsilon_{e\gamma}\over dt} = {4\sigma_{\rm T}a_rT_\gamma^4n_Hx_ek_{\rm B}\over m_ec}(T_\gamma-T_e)\;,
\end{equation}
where $\sigma_{\rm T}$ is the Thomson cross section, $a_r$ is the radiation constant,
$k_{\rm B}$ is the Boltzmann constant, $m_e$ is the electron mass, $T_\gamma$ is the photon
temperature, $T_e$ is the electron temperature, $n_H$ is the nucleon number density
and $x_e$ is the free electron fraction. One further assumes that the electrons preserved
thermal contact with the nucleons, so that $T_e$ was effectively the gas temperature $T_g$.
Using this expression, one may find the redshift $z_{\rm ad}$ at which thermal contact
between the matter and the CMB was broken, which occured when the ratio of the heat content
of the matter to the heat-transfer rate in Equation~(16) exceeded the characteristic
expansion time of the Universe. Beyond this time, matter continued to cool adiabatically,
$T_g\propto (1+z)^2$, at a rate faster than radiation. In $\Lambda$CDM, one finds that
$z_{\rm ad}\approx 150$, and so one expects the gas temperature at $z=17.2$ to have been
\begin{equation}
T_g^{\Lambda{\rm CDM}}(z=17.2)\approx 2.725\;{\rm K}\;{(1+17.2)^2\over 
(1+z_{\rm ad})}\sim 6\;{\rm K}\;.
\end{equation}
But as we have just seen, in order to account for the observed global 21-cm signal, the
gas temperature could not have been higher than $\approx 3.3$ K. Thus, to correctly account
for the observed profile of the 21-cm absorption line, adiabatic cooling in the standard
model would have to have started much earlier, at $z_{\rm ad}\sim 250$. This is the reason
the 21-cm signal measured by EDGES is considered to be an anomaly.

In the $R_{\rm h}=ct$ universe, the temperature at high redshifts (Eq.~8) is lower than
that in the standard model, so adiabatic cooling of the gas would have started even earlier
than in $\Lambda$CDM, due to the much smaller free electron fraction $x_e$. Applying the same
matter-radiation decoupling algorithm described above, one finds that $z_{\rm ad}\gg 150$ in
this model, producing a much colder gas than implied by Equation~(17). This too is inconsistent
with the 21-cm line profile, unless some other mechanism intervened to reheat the gas
at $15\lesssim z\lesssim 20$.

This is where the Pop III stars we discussed in the previous subsection enter the picture.
We pointed out that these early stellar sources not only injected large quantities of dust and
UV photons (which were critical in coupling the spin temperature to the gas temperature via
Eqn.~12) into the IGM, but also irradiated the cosmic background with an X-ray intensity
emitted at the termination shocks of their supernova remnants. As we pointed out,
however, an important caveat with this analysis is that the Pop III star formation rate
required for this to work would have greatly exceeded the predictions in $\Lambda$CDM.
It would have been much more in line with the expansion scenario expected in $R_{\rm h}=ct$,
though the question still remains open regarding which of them is favoured by the data
(see, e.g., ref.~\cite{Steinhardt:2016}).

To produce the amount of dust required to rethermalize the CMB radiation at $z\sim 16$
(see Eqn.~10), we found in \S~3.1 that
\begin{equation}
n_{III}\approx 1.5\times 10^8\;{\rm Mpc}^{-3}
\end{equation}
Pop III stars of average mass $M_{III}=100\;M_\odot$ must have terminated their lives
as supernova explosions. Such stars reach this terminal state as pair-instability
explosions \cite{Heger:2002}, releasing
\begin{equation}
E_{III}\gtrsim 10^{51}\;{\rm erg}
\end{equation}
of energy per event. Within the ejecta, a fraction $\epsilon_e$ of the total supernova
energy is converted into accelerated electrons. The efficiency of this process depends on
the density of the medium surrounding the supernova. For low densities, the shock is weak
and the efficiency is low. Supernova remnants impacting high-density clouds in the local
Universe can have much larger efficiencies. We still do not have a precise understanding
of how the shocks would have evolved in the early IGM, though simulations suggest that the
efficiency might have been near the bottom of the range \cite{Reynolds:2008}, i.e.,
\begin{equation}
\epsilon_e\sim 0.01\;.
\end{equation}

The relativistic electrons produced in this fashion radiated a fraction $f_X$ of their
energy as X-rays, the dominant radiation field responsible for heating the IGM. This
fraction is also subject to uncertainties, e.g., the strength of the magnetic field in
the radiation zone but, based on the modeling of supernova remnants in the local Universe,
one would expect
\begin{equation}
f_X\sim 0.1\;.
\end{equation}

Some attention has already been paid to the fate of this X-radiation in the
high-redshift IGM, once the first generation of Pop~III stars brought an end to
the dark ages. Prior to the reionization of HI gas, UV photons would have been
mostly trapped near their sources, creating localized HII bubbles, but X-rays
would have penetrated to much larger distances. They therefore most likely constituted
the high-energy radiative background throughout the IGM, initiating the ionization of
the neutral gas far from the stars (see, e.g., ref.~\cite{FurlanettoStoever:2010}. Most of
their energy would have been deposited indirectly, starting with photoionization and
subsequently dispersed by the fast electrons released in this process. It is believed
that the electrons energized by X-rays from the Pop~III supernovae \cite{Oh:2001,Venkatesan:2001},
were responsible for heating the IGM before reionization began in earnest
\cite{KuhlenMadau:2005,Furlanetto:2006}. Some have even speculated---well before
the EDGES measurement was known---that this heating and ionization could have important
observational consequences on the highly redshifted 21-cm signal produced by the early
IGM \cite{Furlanetto:2006,KuhlenMadau:2006,Pritchard:2007}.

In their detailed simulations of the energy deposited into the IGM from
the photoionization of neutral atoms by X-rays, ref.~\cite{FurlanettoStoever:2010}
found that ejected electrons with energy $<10.2$ eV could not interact with any
atoms or ions and thus dispersed all of their energy as heat. As the photon
energy increased, however, approaching the X-ray region of the spectrum,
more and more excitation and ionization processes became available and
the fraction of energy deposited as heat decreased. But this trend came
to an end once the number of such process had been saturated, and the
fraction of incoming energy deposited as heat eventually approached a
reasonably constant value at $\sim 1-10$ keV. These authors found that
in highly neutral gas, with $n_{HII}/n_{HI}< 10^{-3}$, approximately
$20\%$ of the electron's energy was deposited as heat, with the remainder
split roughly equally between ionization and excitation. They also found
that these results are insensitive to $n_{HII}/n_{HI}$ at high energies,
because most of the heating comes from secondary electrons with energies
below 10 eV. Since, in addition, only a small fraction of the X-ray photon's
energy is used to photoionize hydrogen in this regime, we may simply write
\begin{equation}
f_{\rm heat}\approx 0.2\;,
\end{equation}
for the fraction of X-ray energy deposited as heat throughout the IGM during
the time (i.e., $15\lesssim z\lesssim 20$) that Pop~III stars ended their
lives as supernova explosions.

Thus, all told, the Pop III X-ray energy converted into IGM heat during the
$15\lesssim z\lesssim 20$ epoch may be estimated as follows:
\begin{eqnarray}
U_{III,\,X}&=&E_{III}\,n_{III}\,\epsilon_e f_X f_{\rm heat}\nonumber\\
&\null&\hskip-0.5in \approx 1.1\times 10^{-18}\left({E_{III}\over 10^{51}\;{\rm erg}}\right)
\left({n_{III}\over 1.5\times 10^8\;{\rm Mpc}^{-3}}\right)\times\nonumber\\
&\null&\hskip 0.1in\left({\epsilon_e\over 0.01}\right)
\left({f_X\over 0.1}\right)\left({f_{\rm heat}\over 0.2}\right)\;{\rm erg}\;{\rm cm}^{-3}.\quad
\end{eqnarray}
As we have seen, adiabatic cooling of the gas in $R_{\rm h}=ct$ would have resulted in
an IGM at $z\sim 20$ much colder than $T_\gamma$, so it is safe to assume that $T_g$
at $z\sim 17.2$ would have been entirely due to the X-ray re-heating implied by
Equation~(23). And if we were to put
\begin{equation}
U_{III,\,X}={3\over 2}n_HkT_g
\end{equation}
(given that both the electron and $n_I$ densities were negligible compared to
$n_H$), we would find that
\begin{eqnarray}
T_g&\approx& 3.4\left({E_{III}\over 10^{51}\;{\rm erg}}\right)
\left({n_{III}\over 1.5\times 10^8\;{\rm Mpc}^{-3}}\right)\times\nonumber\\
&\null&\hskip 0.1in\left({\epsilon_e\over 0.01}\right)
\left({f_X\over 0.1}\right)\left({f_{\rm heat}\over 0.2}\right)\;{\rm K}\;.\quad
\end{eqnarray}

One would like to think that the remarkable consistency between the spin
temperature ($T_{\rm s}\approx 3.3$ K) measured by the EDGES collaboration and the gas
temperature we have just estimated ($T_g\approx 3.4$ K) is strong evidence in favour of
this model. Of course, this is almost certainly partially a coincidence, given that every
parameter in Equation~(25) contributing to this outcome was assigned its fiducial value.
On the other hand, the fact that this analysis produced this level of consistency without
any optimization of the physical inputs does suggest that the scenario we have explored
here is at least viable. None of the estimates had to be `stretched' to unlikely values,
and all were internally consistent with the basic premise that the CMB we see was produced
by dust rethermalization of the primordial radiation field during the Pop III
star-formation era. The principal caveat with this analysis concerns
the Pop III star formation rate, which differs by several orders of magnitude between
the $\Lambda$CDM and $R_{\rm h}=ct$ cosmologies. The picture we have painted in this
paper holds together self-consistently as long as the $R_{\rm h}=ct$ rate is correct,
but would break down if future observations reveal that the large-scale structure formation
rate at $z\gtrsim 10$ was more in line with the predictions of the standard model.

Once the global 21-cm signal is confirmed by upcoming observations, a more
detailed simulation than that attempted here ought to be carried out, highlighting the
evolution of the physical state of the IGM from $z\sim 20$ to $z\sim 15$. The EDGES
absorption line was seen throughout this region, so a natural question to address is 
whether the brightness temperature calculated with Equation~(15) is consistent with the
radiation ($T_\gamma$) and spin ($T_{\rm s}$) temperatures we have estimated thus
far. The answer appears to be yes, given that $T_\gamma$ changes slowly at $z\gtrsim 16$,
evolving from the free-streaming relation in Equation~(7) to the expression in 
Equation~(8). So, for example, while $T_\gamma\approx 49.6$ K at $z=17.2$, it
increases only slightly to $\approx 53.3$ K by $z=20$, implying that
$\delta T_b\approx -495$ mK at $z=17.2$ and $-574$ mK at $z=20$. The physical
conditions therefore appear to be quite similar throughout the $15\lesssim z
\lesssim 20$ epoch though, in reality, there must have been an evolution in
the heating rate of the IGM gas as the number of Pop III SNe ramped up. A
more detailed calculation, with a more realistic Pop III star formation rate
throughout this period, should show whether the inferred physical state of the 
IGM is truly consistent with the observed 21-cm line profile.

\begin{figure}
\vskip 0.1in
\centering
\includegraphics[width=0.9\linewidth]{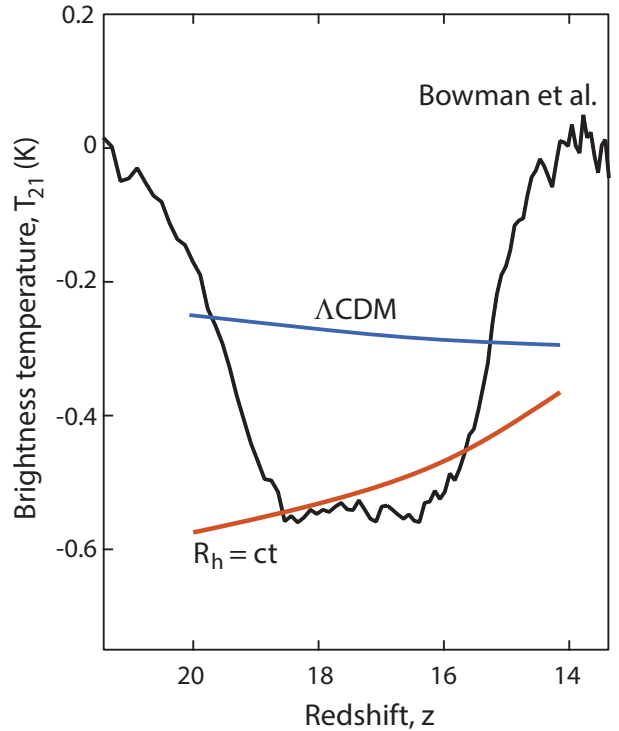}
\caption{Best-fitting 21-cm absorption profile (solid black) for the brightness 
temperature $T_{21}$ corresponding to the hardware and analysis configuration with
the highest signal-to-noise ratio (equal to 52), versus redshift, reported by
Bowman et al. \cite{Bowman:2018}, compared with the predicted brightness temperature
in the $\Lambda$CDM (solid blue) and $R_{\rm h}=ct$ (solid red) cosmologies.
\label{fig1}}
\end{figure}

Nevertheless, even at this level of approximation, the brightness temperature
of the 21-cm signal predicted by the $R_{\rm h}=ct$ universe appears to be more consistent
with the EDGES observation than that calculated using the standard model (see fig.~\ref{fig1}).
As noted earlier, the limitation with our simplified approach in this paper is that the physical 
conditions within the redshift range $15\lesssim z\lesssim 20$ are estimated more or less as 
averages, so we are not yet in a position to show a true time-evolution across this transition 
region. The brightness temperature (Eqn.~15), however, does trace the evolution in CMB temperature
according to Equation~(7) in $\Lambda$CDM, and according to a transition from Equation~(8)
to (7) in $R_{\rm h}=ct$, so we may at least gauge how well the 21-cm line profile matches
the CMB temperature in these two models within this redshift range. As one may see in
figure~\ref{fig1}, the correspondence is much better for $R_{\rm h}=ct$ than for $\Lambda$CDM, 
largely due to two principal effects: (i) the difference in spin temperature predicted by these 
two cosmologies ($T_{\rm s}\approx 3.4$ K in the former, versus $\approx 6$ K in the latter);
and (ii) the somewhat different redshift dependence of the CMB temperature at $z\gtrsim 16$,
given by Equation~(7) in $\Lambda$CDM and Equation~(8) in $R_{\rm h}=ct$.

\section{Discussion and Conclusion}
There is no question that the debate between a recombination versus dust origin for
the CMB has shifted strongly in favour of the former following the dramatic discoveries
by COBE, WMAP and {\it Planck} over the past two decades. Yet this issue has slowly
resurfaced in the face of growing tension between the predictions of the standard model
and the ever improving precision of cosmological measurements. A good example is the
Hubble constant $H_0$, which characterizes the current expansion rate of the Universe
and determines its absolute distance scale. The accuracy of measuring $H_0$ has been
significantly improved recently, but its value ($67.4\pm0.5$ km $\rm s^{-1}$ $\rm Mpc^{-1}$;
ref.~\cite{Planck:2016} inferred from the CMB observations in the context of flat
$\Lambda$CDM disagrees with that based on local Type Ia SNe calibrated by the Cepheid
distance ladder ($74.03\pm1.42$ km $\rm s^{-1}$ $\rm Mpc^{-1}$; ref.~\cite{Riess:2019})
at a $4.4\sigma$ level of significance.

A previously completed thorough analysis of the evidence in favour of recombination
\cite{Melia:2020b} demonstrated that a dust scenario for the origin of the CMB is still
viable, though not in the context of $\Lambda$CDM. Instead, the observed characteristics
of the microwave background, such as its distribution of anisotropies and the implied
acoustic scale, could be consistent with the alternative FLRW cosmology known as
the $R_{\rm h}=ct$ universe, which is effectively $\Lambda$CDM, though with an
additional constraint on its overall equation-of-state derived from the zero
active mass condition in general relativity \cite{Melia:2020a}. The work we have reported
in this paper has a significant impact on this discussion because it represents one
of the earliest tests of the dust model based on a very different kind of observation.

The observed profile of the global 21-cm signal associated with the CMB is largely
consistent with the expectations of $\Lambda$CDM, except that its amplitude is more than
a factor of two greater than the largest predictions. Like the disparity with $H_0$,
this anomaly represents a statistically significant indication that the standard model
may not be quite right. It may need an infusion of new physics or, at worst, a complete
replacement. The $R_{\rm h}=ct$ universe is a compromise, because it actually does
not replace any of the essential features of the standard model (other than the
inflationary phase, which itself appears to be problematic; ref.~\cite{Melia:2013}), but only adds
the requirement that $\rho+3p=0$ at all times. From a theoretical standpoint, the
Local Flatness Theorem in general relativity \cite{Melia:2019a} leaves no doubt that
the use of FLRW is valid only with the inclusion of this equation-of-state. And its
observational support continues to grow with the completion of each new comparative
test (see, e.g., Table~2 in ref.~\cite{Melia:2018a}). The fact that $R_{\rm h}=ct$ presents a
viable mechanism by which the measured 21-cm signal could have formed at
$15\lesssim z\lesssim 20$ constitutes additional observational support for this
model.

If confirmed, the EDGES detection \cite{Bowman:2018} is critically important to
astrophysics in general, perhaps even to particle physics, because it opens up a
window on the early phase of cosmic structure formation, and provides us with the
physical state of the cosmic environment at the start of the epoch of reionization.
As we have seen in this paper, the requirements for the rethermalization of the
CMB by dust at $z\sim 16$ in the context of $R_{\rm h}=ct$ seemlessly weave together
a narrative in which the onset of Pop III star formation at $15\lesssim z\lesssim 20$
self-consistently initiated a re-heating of the IGM, an increase in metallicity
and the concomitant injection of dust into the background medium. A single
population of stars produced all of the physical attributes needed to account
for the CMB spectrum, the apparent frequency dependence of its distribution of
anisotropies, and an equilibrium IGM gas (and spin) temperature reflected in
the global 21-cm profile produced during this epoch. This represents a comprehensive
set of developments fulfilling the cosmic transition away from the dark ages and
into the subsequent era of reionization and large-scale formation of structure.

Fortunately, it appears that we shall not have to wait very long to see a confirmation
of the EDGES measurement. Several other similar experiments are underway, including
the Large-Aperture Experiment to Detect the Dark Ages (LEDA \cite{Bernardi:2016};
the Sonda Cosmol\'ogica de las Islas para la Detecci\'on de Hidr\'ogeno Neutro
(SCI-HI \cite{Voytek:2014}); and the Shaped Antenna measurement of the background Radio Spectrum 2
(SARAS2 \cite{Singh:2017}). Farther afield, the observation of the 21-cm line
should be significantly enhanced by the use of interferometric arrays, such as
the Hydrogen Epoch of Reionization Array (HERA \cite{DeBoer:2017}); and the
Square Kilometre Array (SKA; https://www.skatelescope.org), among others. When
constructed, the SKA Low-Frequency Aperture Array will detect the power spectrum
associated with the EDGES absorption profile, and should also be able to image
the 21-cm signal, providing more fertile ground for testing the scenario we have
explored in this paper, in which the physical conditions producing the 21-cm
absorption line are inextricably linked to the requirements for rethermalizing
the CMB at $z\sim 16$, just prior to the epoch of reionization.

{\acknowledgement
I am very grateful to the anonymous referee for their exceptional
review of this paper and very helpful suggestions that have led to an improved
presentation of the results. I am also grateful to Amherst College for its support 
through a John Woodruff Simpson Lectureship, and Purple Mountain Observatory in 
Nanjing, China, for its hospitality while part of this work was being carried out.
\endacknowledgement}

%=====================================================
%
%                           BIBLIOGRAPHY
%
%=====================================================

%=====================================================

\begin{thebibliography}{99}
\bibitem{Bowman:2018} J. D. Bowman, A.E.E. Rogers, R. A. Monsalve, T. J. Mozdzen \& N. Mahesh,
Nature {\bf 555} (2018) 67
\bibitem{Bernardi:2015} G. Bernardi, M. McQuinn \& L. J. Greenhill, ApJ {\bf 799} (2015) 90
\bibitem{Pritchard:2014} J. Pritchard, K. Ichiki, A. Mesinger, R. B. Metcalf, 
A. Pourtsidou, M. Santos, F. B. Abdalla et al., Proceedings of Advancing Astrophysics with 
the Square Kilometre Array (AASKA14). 9-13 June, 2014. Giardini Naxos, Italy. Online at 
http://pos.sissa.it/cgi-bin/reader/conf.cgi?confid=215 (2014) id 12
\bibitem{MeliaShevchuk:2012} F. Melia \& A. Shevchuk, MNRAS {\bf 419} (2012) 2579
\bibitem{Melia:2020a} F. Melia, {\it The Cosmic Spacetime} (New York: Taylor \& Francis) (2020)
\bibitem{Melia:2019a} F. Melia, Annals of Physics {\bf 411} (2019) 167997
\bibitem{Melia:2016} F. Melia, Front. Phys. {\bf 11} (2016) 119801
\bibitem{Melia:2017} F. Melia, Front. Phys. {\bf 12} (2017) 129802
\bibitem{Melia:2018a} F. Melia, MNRAS {\bf 481} (2018) 4855
\bibitem{Melia:2018b} F. Melia, Am. J. Phys. {\bf 86} (2018) 585
\bibitem{MeliaMaier:2013} F. Melia \& R. S. Maier, MNRAS {\bf 432} (2013) 2669
\bibitem{Melia:2020b} F. Melia, EPJ Plus {\bf 135} (2020) 511
\bibitem{Planck:2016} Planck Collaboration, A\&A {\bf 594} (2016) A15
\bibitem{PeeblesYu:1970} P.J.E. Peebles \& J. T. Yu, ApJ {\bf 162} (1970) 815
\bibitem{White:1994} M. White \& J. Silk, ARAA {\bf 32} (1994) 319
\bibitem{Hu:1995} W. Hu \& N. Sugiyama, ApJ {\bf 444} (1995) 489
\bibitem{Seo:2005} H.-J. Seo \& D. J. Eisenstein, ApJ {\bf 633} (2005) 575
\bibitem{Meiksin:1999} A. Meiksin, M. White \& J. A. Peacock, MNRAS {\bf 304} (1999) 851
\bibitem{Jeong:2006} D. Jeong \& E. Komatsu, ApJ {\bf 651} (2006) 619
\bibitem{Crocce:2006} M. Crocce \& R. Scoccimarro, Phys. Rev. D {\bf 73} (2006) 063520
\bibitem{Eisenstein:2007} D. J. Eisenstein, H.-J. Seo \& M. White, ApJ {\bf 664} (2007) 660
\bibitem{Nishimichi:2007} T. Nishimichi, H. Ohmuro, M. Nakamichi et al., PASJ {\bf 59} (2007) 1049
\bibitem{Seo:2010} H.-J. Seo et al., ApJ {\bf 720} (2010) 1650
\bibitem{Font-Ribera:2014} A. Font-Ribera et al., JCAP {\bf 5} (2014) 27
\bibitem{Delubac:2015} T. Delubac, J. E. Bautista, N. G. Busca, J. Rich, D. Kirkby, S. Bailey,
A. Font-Ribera et al., A\&A {\bf 574} (2015) A59
\bibitem{Alam:2017} A. Alam, M. Ata, S. Bailey, F. Beutler, D. Bizyaev, J. A. Blazek, A. S. Bolton
et al., MNRAS {\bf 470} (2017) 2617
\bibitem{MeliaLopez:2017} F. Melia \& M. L\'opez-Corredoira, IJMP-D {\bf 26} (2017) 1750055
\bibitem{MeliaLopez:2018} F. Melia \& M. L\'opez-Corredoira, A\&A {\bf 610} (2018) A87
\bibitem{Copi:2015} C. J. Copi, D. Huterer, D. J. Schwarz \& G. D. Starkman, MNRAS {\bf 451} (2015) 2978
\bibitem{Mather:1990} J. C. Mather, E. S. Cheng, R. E. Jr. Eplee, R. B. Isaacman, S. S. Meyer, 
R. A. Shafer, R. Weiss et al., ApJL {\bf 354} (1990) L37
\bibitem{Rubino-Martin:2006} J. A. Rubino-Martin, J. Chluba \& R. A. Sunyaev, MNRAS {\bf 371} (2006) 1939
\bibitem{Rubino-Martin:2008} J. A. Rubino-Martin, J. Chluba \& R. A. Sunyaev, A\&A {\bf 485} (2008) 377
\bibitem{Rees:1978} M. J. Rees, Nature {\bf 275} (1978) 35
\bibitem{Rowan:1979} M. Rowan-Robinson, J. Negroponte \& J. Silk, Nature {\bf 281} (1979) 635
\bibitem{Wright:1982} E. L. Wright, ApJ {\bf 255} (1982) 401
\bibitem{Planck:2018} Planck Collaboration, A\&A in press (arXiv:1906.02552) (2018)
\bibitem{Weingartner:2001} J. C. Weingartner \& B. T. Draine, ApJ {\bf 548} (2001) 296
\bibitem{Draine:2001} B. T. Draine \& A. Li, ApJ {\bf 551} (2001) 807
\bibitem{Bromm:2004} V. Bromm \& R. B. Larson, ARA\&A {\bf 42} (2004) 79
\bibitem{Glover:2004} S.C.O. Glover, Space Sci Review {\bf 117} (2004) 445
\bibitem{Heger:2003} A. Heger, C. L. Fryer, S. E. Woosley, N. Langer \& D. H. Hartmann, 
ApJ {\bf 591} (2003) 288
\bibitem{Whalen:2008} D. Whalen, B. van Veelen, B. W. O'Shea \& M. L. Norman, ApJ {\bf 682} (2008) 49
\bibitem{Heger:2002} A. Heger \& S. E. Woosley, ApJ {\bf 567} (2002) 532
\bibitem{Xu:2016a} H. Xu, M. Norman, B. W. O'Shea \& J. H. Wise, ApJ {\bf 823} (2016) 140
\bibitem{Xu:2016b} H. Xu, M. Norman, B. W. O'Shea \& J. H. Wise, ApJ {\bf 833} (2016) 84
\bibitem{Mebane:2018} R. H. Mebane, J. Mirocha \& S. R. Furlanetto, MNRAS {\bf 479} (2018) 4544
\bibitem{Visbal:2018} E. Visbal, Z. Haiman \& G. L. Bryan, MNRAS {\bf 475} (2018) 5246
\bibitem{Jaacks:2019} J. Jaacks, S. L. Finkelstein \& V. Bromm, MNRAS {\bf 488} (2019) 2202
\bibitem{Sarmento:2019} R. Sarmento, E. Scannapieco \& B. C{\^o}t{\'e}, ApJ {\bf 871} (2019) 206
\bibitem{Melia:2019b} F. Melia, EPJ-C {\bf 79} (2019) 455
\bibitem{Melia:2014} F. Melia, AJ {\bf 147} (2014) 120
\bibitem{Yennapureddy:2018} K. M. Yennapureddy \& F. Melia, PDU {\bf 20} (2018) 50
\bibitem{Yennapureddy:2019} K. M. Yennapureddy \& F. Melia, EPJ-C {\bf 79} (2019) 571
\bibitem{Yennapureddy:2020} K. M. Yennapureddy \& F. Melia, PDU in press (2020) 
\bibitem{Steinhardt:2016} C. L. Steinhardt, P. Capak, D. Masters \& J. S. Speagle, 
ApJ {\bf 824} (2016) 21
\bibitem{Melia:2015} F. Melia \& T. M. McClintock, Proc. R. Soc. A {\bf 471} (2015) 20150449
\bibitem{Pritchard:2012} J. R. Pritchard \& A. Loeb, Rep. Prog. Phys. {\bf 75} (2012) 086901
\bibitem{Wouthuysen:1952} S. A. Wouthuysen, AJ {\bf 57} (1952) 31
\bibitem{Hirata:2006} C. M. Hirata, MNRAS {\bf 367} (2006) 259
\bibitem{Zygelman:2005} B. Zygelman, ApJ {\bf 622} (2005) 1356
\bibitem{Zaldarriaga:2004} M. Zaldarriaga, S. R. Furlanetto \& L. Hernquist, ApJ {\bf 608} (2004) 622
\bibitem{Peebles:1968} P.J.E. Peebles, ApJ {\bf 153} (1968) 1
\bibitem{Weymann:1965} R. Weymann, Phys. Fluids {\bf 8} (1965) 2112
\bibitem{Reynolds:2008} J. P. Reynolds, ARAA {\bf 46} (2008) 89
\bibitem{FurlanettoStoever:2010} S. R. Furlanetto \& S. J. Stoever, MNRAS {\bf 404} (2010) 1869
\bibitem{Oh:2001} S. P. Oh, ApJ {\bf 553} (2001) 499
\bibitem{Venkatesan:2001} A. Venkatesan, M. L. Girous \& J. J. Shull, ApJ {\bf 563} (2001) 1
\bibitem{KuhlenMadau:2005} M. Kuhlen \& P. Madau, MNRAS {\bf 363} (2005) 1069
\bibitem{Furlanetto:2006} S. R. Furlanetto, MNRAS {\bf 371} (2006) 867
\bibitem{KuhlenMadau:2006} M. Kuhlen, P. Madau \& R. Montgomery, ApJL {\bf 637} (2006) L1
\bibitem{Pritchard:2007} J. J. Pritchard \& S. R. Furlanetto, MNRAS {\bf 376} (2007) 1680
\bibitem{Riess:2019} A. G. Riess, S. Casertano, W. Yuan, L. M. Macri \& D. Scolnic, 
ApJ {\bf 876} (2019) 85
\bibitem{Melia:2013} F. Melia, A\&A {\bf 553} (2013) A76
\bibitem{Bernardi:2016} G. Bernardi et al., MNRAS {\bf 461} (2016) 2847
\bibitem{Voytek:2014} T. C. Voytek, A. Natarajan, J. Jauregui Garcia, J. B. Peterson \& 
O. L\'opez-Cruz, ApJL {\bf 782} (2014) L9
\bibitem{Singh:2017} S. Singh et al., ApJL {\bf 845} (2017) L12
\bibitem{DeBoer:2017} D. R. DeBoer et al., PASP {\bf 129} (2017) 045001
\end{thebibliography}
\end{document}